\documentclass[iop,tighten,twocolumn,preprintnumbers]{emulateapj}
\usepackage{lineno}
\modulolinenumbers[5]
\usepackage{amsmath,bbold,amssymb,epsfig,bm,mathrsfs,feynmp}
\usepackage{color,slashed,nicefrac,exscale,multirow,soul,float}
\usepackage{times,txfonts,xcolor,graphicx,gensymb,tikz}
\usepackage[normalem]{ulem}
\usepackage[breaklinks,colorlinks,citecolor=blue]{hyperref}
\usepackage{eso-pic}
\usepackage{lineno}
\definecolor{red}{rgb}{0.8,0,0}
\definecolor{violet}{rgb}{0.4,0,0.4}
\definecolor{green}{rgb}{0,0.5,0.0}
\definecolor{navy}{rgb}{0.0,0.0,0.6}
\definecolor{orange}{rgb}{0.8,0.2,0.0}

\newcommand{\blue}[1]{\textcolor[rgb]{0.00,0.00,1.00}{#1}}

\newcommand{\bea}{\begin{eqnarray}}
\newcommand{\eea}{\end{eqnarray}}
\newcommand{\ep}{\varepsilon}
\begin{document}
\title{Hybrid star models in the light of new multi-messenger data\vspace{-1.6cm}}
\author{Jia Jie Li$^1$,
  Armen Sedrakian$^{2,3}$,
  Mark Alford$^4$}
\affiliation{
$^1$School of Physical Science and Technology, 
    Southwest University, Chongqing 400715, China;
    \blue{jiajieli@swu.edu.cn}\\
$^2$Frankfurt Institute for Advanced Studies,
    D-60438 Frankfurt am Main, Germany;
    \blue{sedrakian@fias.uni-frankfure.de}\\
$^3$Institute of Theoretical Physics,
    University of Wroclaw, 50-204 Wroclaw, Poland\\    
$^4$Department of Physics, 
    Washington University, St.~Louis, MO 63130, USA;
   \blue{alford@physics.wustl.edu}\\
}
\begin{abstract}
Recent astrophysical mass inferences of compact stars HESS J1731-347 
and PSR J0952-0607, with extremely small and large masses respectively, 
as well as the measurement of the neutron skin of Ca in the CREX experiment 
challenge and constrain the models of dense matter. We examine the concept 
of hybrid stars - objects containing quark cores surrounded by nucleonic 
envelopes - as models that account for these new data along with other 
inferences. We employ a family of 81 nucleonic equations of state (EoSs) 
with variable skewness and slope of symmetry energy at saturation density 
and a constant speed-of-sound EoS for quark matter. For each nucleonic EoS, 
a family of hybrid EoSs is generated by varying the transition density, the 
energy jump, and the speed of sound. These models are tested against the 
data from GW170817 and J1731-347, which favor low-density soft EoS and 
J0592-0607 and J0740+6620, which require high-density stiff EoS. 
The addition of J0592-0607's mass measurement to the constraints has 
no significant impact on the parameter space of the admissible EoS, but 
allows us to explore the potential effect of pulsars more massive than 
J0740+6620, if such exists.
We then examine the occurrence of twin configurations and quantify the ranges 
of masses and radii that they can possess. It is shown that including J1731-347 
data favors EoSs that predict low-mass twins with $M \lesssim 1.3\,M_{\odot}$ 
that can be realized if the deconfinement transition density is low. If combined 
with large speed of sound in quark matter such models allow for maximum masses 
of hybrid stars in $2.0$--$2.6\,M_{\odot}$.
\end{abstract}

\section{Introduction}
\label{sec:Intro}
Although the concepts of quark stars and hybrid stars containing a quark 
core enclosed in a nucleonic envelope were proposed  long ago and have been 
intensively studied over the following decades 
(for reviews, see~\cite{Alford:2007,Anglani:2013,Baym:2018,Sedrakian:2023part}, 
they remain at the forefront of the exploration of superdense matter and 
compact stars (CSs). They may arise through the onset of a first-order 
phase transition between the nucleonic and quark phases, in which case a 
new branch of stellar equilibria populated by stable hybrid stars arise. 
One intriguing aspect of hybrid stars is the existence of twin configurations, 
i.e, two stars that possess the same mass but different radii, whereby the 
larger star consists entirely of nucleonic matter, while the more compact 
star is a hybrid star~\citep{Glendenning:2000,Zdunik:2013,Benic:2015,Alford:2017,Alvarez-Castillo:2019,Blaschke:2020,Christian:2022,Lijj:2023b}.
The study of hybrid stars and their twin configurations continues to
be an active area of research, with ongoing efforts to explore their
properties and to investigate potential observational signatures that
could distinguish them from hadronic CSs~\citep[see][and references
therein]{Alford:2005,Bonanno:2012,Masuda:2013,Klahn:2013,Christian:2019,Malfatti:2020,Lijj:2020a,Tan:2020,Tan:2022,Brodie:2023,Tsaloukidis:2023}.

\begin{figure*}[tb]
\centering
\includegraphics[width = 0.96\textwidth]{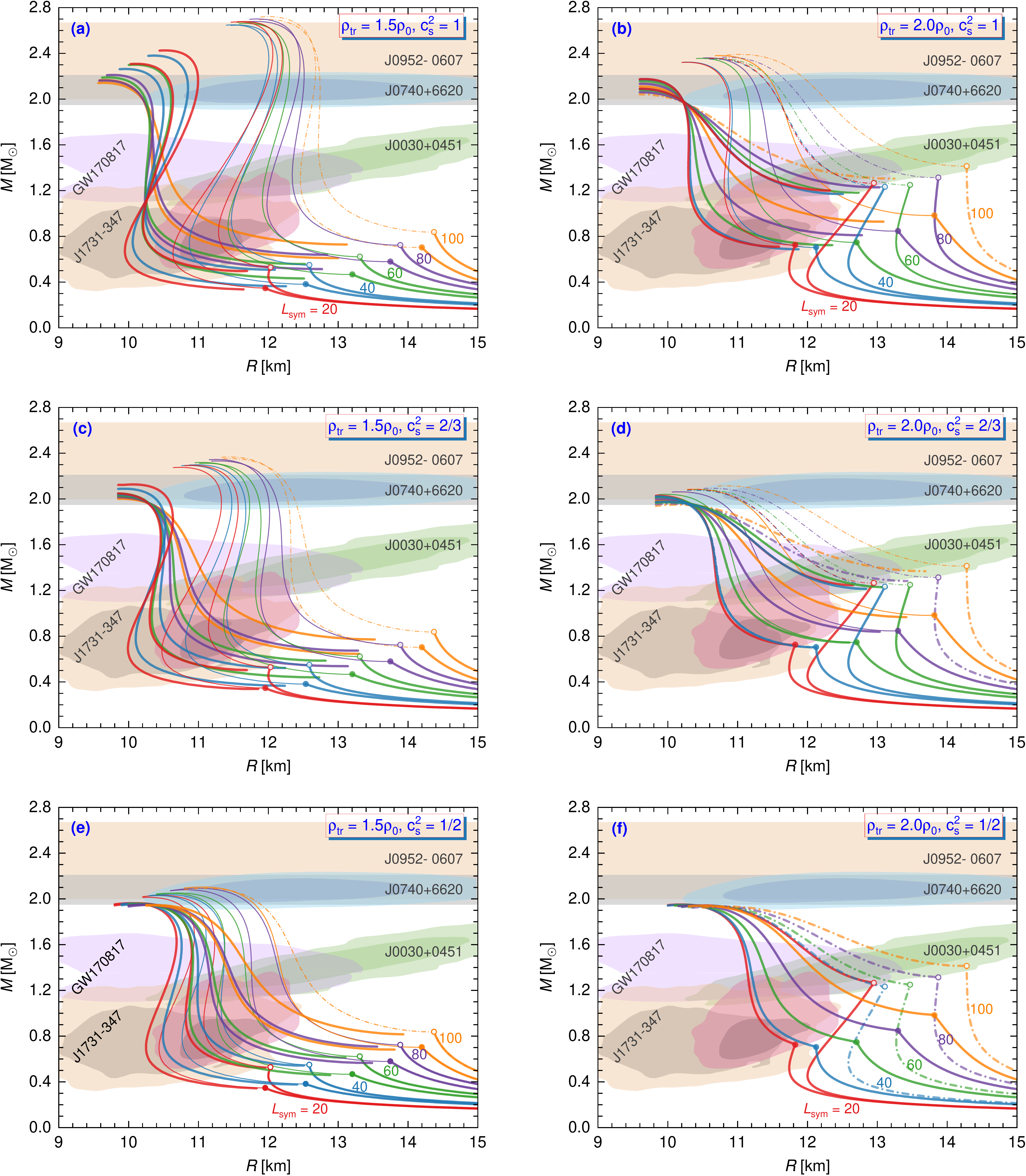}
\caption{ Mass-radius relations for hybrid EoSs with
  $\rho_{\rm tr} = 1.5\,\rho_0$ and $2.0\,\rho_0$, for
  $c^2_s = 1$, $2/3$ and $1/2$ in the quark phase. In each panel,
  the hybrid EoSs are built upon nucleonic models grouped by
  $L_{\rm sym} = 20$, 40, 60, 80 and 100\,MeV; for each, $Q_{\rm sat}$
  is set to -600 and 1000\,MeV. The circle on each curve denotes the
  configuration with central density $\rho_{\rm tr}$; lines ending 
  with a filled circle are for the isoscalar-soft nucleonic model 
  with $Q_{\rm sat}=-600$\,MeV and lines ending with an empty
  circle are for the isoscalar-stiff one with $Q_{\rm sat}=1000$\,MeV. 
  Observational constraints from multi-messenger astronomy are
  shown for comparison. 
  The two overlapping regions for HESS J1731-347~\citep{Doroshenko:2022} 
  coincide with areas bounded by solid and dashed lines 
  in their Figure~1. For each nucleonic EoS, the thick line shows the 
  model with a maximum energy jump $\Delta\ep$ that yields hybrid branch 
  passing through the 95\% lower limit for the mass-radius constraint of 
  PSR J0740+6620 or J0030+0451, while the thin line shows the model with 
  a critical value of $\Delta\ep$ beyond which  a disconnected 
  mass-radius curve arises. Solid lines represent models that are satisfying 
  all constraints, while dash-dotted lines show those failing for J1731-347. 
  Twin configurations arise for reasonable ranges of model parameters
  and for $c^2_s = 1$ and $2/3$.
  }
\label{fig:MR_diagram1}
\end{figure*}

In recent years, observations of CSs have reached unprecedented levels
of precision, providing valuable constraints on the EoS of stellar
matter and its composition. Among these constraints, the most
interesting ones are depicted in Figure~\ref{fig:MR_diagram1}. These
include the following:\\
(a)~The most massive pulsar PSR J0740+6620~\citep{Fonseca:2021}, which
has pushed the maximum mass of CSs to $2.0\,M_{\odot}$.\\
(b)~The simultaneous mass and radius measurements for PSR
J0030+0451~\citep{NICER:2019a,NICER:2019b} and
J0740+6620~\citep{NICER:2021a,NICER:2021b} extracted from X-ray
observations using the NICER instrument. \\
(c)~The detection of gravitational waves from the binary CS merger
event GW170817~\citep{LIGO-Virgo:2019}, which provides us with an
estimate of the tidal deformability of the binary that can further
yield constraints on the radii of the individual components. 
Theoretical modeling of GW170817 led to {\it upper bound} 
on the mass of a static CS 
$M\le 2.3\,M_{\odot}$~\citep{Rezzolla:2018,Shibata:2019,Khadkikar:2021},
but it is still uncertain and will not be considered below.

In addition to these well-established constraints, some
new results which are awaiting confirmation, 
have been reported very recently: \\
(d)~The companion of the ``black widow" pulsar PSR J0952-0607 (the
second-fastest known spinning CS with a frequency of 707\,Hz) has been
detected in the Milky Way, with a mass estimate of
$M = 2.35^{+0.17}_{-0.17}\,M_{\odot}$ (68\% credible interval)~\citep{Romani:2022}. \\
(e)~The X-ray spectrum of the central compact object within the 
supernova remnant HESS J1731-347 was modeled and in combination with 
the distance estimates from Gaia observations yielded a very low mass 
of $M = 0.77^{+0.20}_{-0.17}\,M_{\odot}$ and a relatively small radius 
of $R = 10.4^{+0.86}_{-0.78}$\,km (68\% credible interval)~\citep{Doroshenko:2022}. 
This modeling substantially revised the previous inference 
of~\cite{Klochkov2015} which was based on an alternate value of the 
distance 3.2\,kpc to this object. \cite{Doroshenko:2022} show six 
different ellipses which correspond to a step-wise increase in the 
realism and improvement of their modeling via: 
(a) inclusion of data below 1\,keV, 
(b) replacement of the {\it wabs} simple absorption model developed 
by~\cite{Morrison1983} by the {\it tbabs} model of~\cite{Wilms2000} that 
provides a more accurate description of the absorption process, 
(c) accounting for the scattering component, 
(d) fixing the distance to the smaller and more realistic value of 
2.5\,kpc, and finally (e) using distance priors from the Gaia estimate 
of the optical companion of HESS J1731-347. As seen from Extended Data 
Figure~1 in~\cite{Doroshenko:2022} the steps (d) and (e) shift the 
mass-radius ellipse to smaller values than previously estimated. 
The last two ellipses are then considered as the most realistic 
because they include the most sophisticated physical modeling and 
the most accurate information on the distance to the object. These 
authors also provide in their Table 1 best fits using five different
atmosphere models. The three models of single-temperature atmospheres 
provide quantitatively similar values, which are adopted as the most 
reliable case. The other two-temperature models were considered having 
fixed $1.4\,M_{\odot}$ mass and 12\,km radius. The temperatures of 
the two bright spots and the rest surface as well as the size of the 
spots were fitting parameters. While it was shown that neutron stars 
with non-uniform surfaces and the above-quoted conventional mass and 
radius values can also describe the observed spectra, the absence of 
pulsation (expected in this case) poses a serious problem for such 
two-temperature models (V.~Suleimanov, private communication). 

Given these new inferences, it is worthwhile to take a fresh look at 
the concept of a hybrid star by allowing for a broad range of admissible 
parameter space within models that have been tested on a more limited 
set of observations. In the present work, we implement such a program 
for the construction of hybrid EoSs and apply filtering to extract the 
range of EoS that is consistent with the current data. Our further focus 
is on twin stars and their compatibility with the aforementioned 
multi-messenger information after the inclusion of extreme domains 
of masses in our analysis.

\section{Hybrid models}
\label{sec:Hybrid}
To describe low-density matter, we use a covariant density functional 
of nuclear matter in which baryons are coupled to mesons with 
density-dependent couplings~\citep{Typel:1999,Oertel:2017,Sedrakian:2023ppnp}. 
The nuclear matter properties are conveniently assessed using the coefficients 
of Taylor expansion of energy density with respect to the baryon density and 
isospin asymmetry at the saturation density $\rho_0$. The low-order coefficients, 
the saturation energy $E_{\rm sat}$, compressibility $K_{\rm sat}$, and 
symmetry energy $J_{\rm sym}$ are either strongly constrained or have no noticeable 
impact on the gross properties (e.g., mass, radius, deformability, etc.) 
of CSs~\citep{Margueron:2018,Lijj:2019a}.

\begin{figure*}[tb]
\centering
\includegraphics[width = 0.96\textwidth]{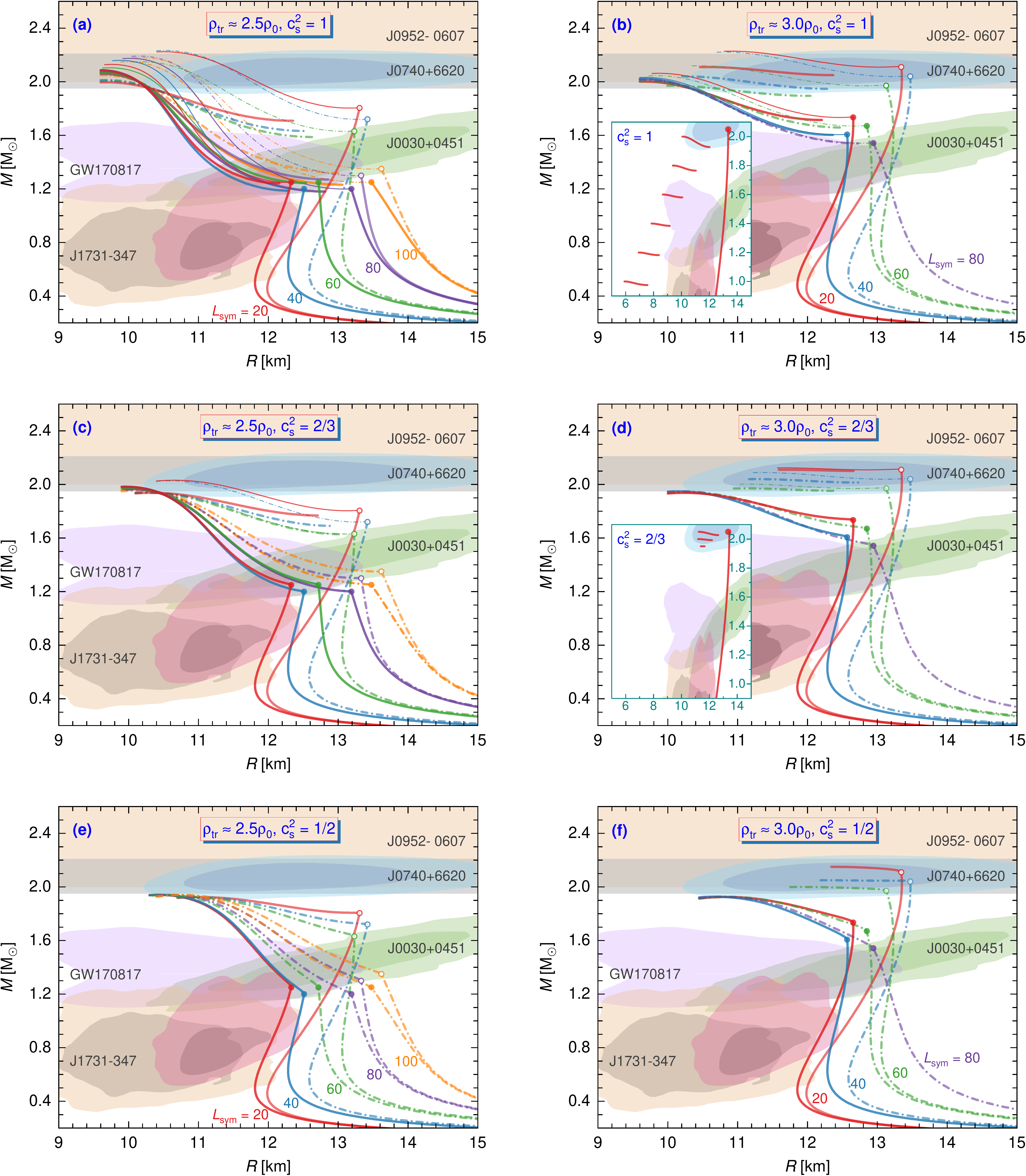}
\caption{ Mass-radius relations for hybrid EoSs with
  $\rho_{\rm tr} \simeq 2.5\,\rho_0$ and $3.0\,\rho_0$, for $c^2_s = 1$,
  $2/3$ and $1/2$ in the quark phase. In each panel, the hybrid EoSs
  are constructed from nucleonic models grouped by
  $L_{\rm sym} = 20$, 40, 60, 80 and 100\,MeV; for each, $Q_{\rm sat}$
  is chosen such that the resultant mass-radius curve passes through
  the 95\% confidence region for HESS J1731-347 (solid lines) or
  GW170817 (dash-dotted lines), respectively. The circle on each curve
  denotes the configurations with central density $\rho_{\rm tr}$. 
  For each nucleonic EoS, the thick line shows the model with a maximum 
  $\Delta\ep$ that yields a hybrid branch passing through the 95\% lower 
  limit for the mass-radius constrain of PSR J0740+6620, while the thin 
  line shows the model with a connected mass-radius curve with a critical 
  value of $\Delta\ep$ set by the constraint for J0740+6620 or appearance 
  of twin configurations. Solid lines represent models satisfying all 
  constraints, while dash-dotted lines show those failing for J1731-347. 
  The inset illustrates hybrid EoSs built upon an isoscalar-stiff nucleonic 
  model for the case $M^{\rm Q}_{\rm max} \leq M^{\rm H}_{\rm max}$ which 
  yields ultra-compact configurations for the case $c^2_s = 1$. 
  }
\label{fig:MR_diagram2}
\end{figure*}

Here we adopt a family of nucleonic EoSs consisting of 81 EoSs that
share the same low-order coefficients as the original DDME2
parameterization~\citep{Lalazissis:2005}, but have alternative
density-dependence of the couplings, thus, resulting in different
high-order coefficients~\citep{Lijj:2023c}.  
These coefficients and their variations explored in this work 
are summarized in Table~\ref{tab:Nuclear_matter}. It is worthwhile 
mentioning that in the calibration of model parameters, we fixed the 
symmetry energy at the crossing density 0.110\,fm$^{-3}$ (which is 
tightly constrained from finite nuclei properties) as that of DDME2, 
the symmetry energy and its slope coefficient at saturation density 
are thus correlated~\citep{Lijj:2023c}. Specifically, the skewness
coefficient for symmetric matter (characterizing the isoscalar sector)
varies in the range $-600 \le Q_{\rm sat} \le 1000$\,MeV and the slope
parameter of the symmetry energy (characterizing the isovector sector)
varies in the range $20 \le L_{\rm sym} \le 100$\,MeV. Note that
values of $Q_{\rm sat}$ and $L_{\rm sym}$ control, respectively, the
high- and intermediate-density behaviors of the nucleonic EoS, thereby
influencing the maximum mass of static nucleonic stars and the radius
of the canonical-mass star~\citep{Margueron:2018,Lijj:2019a}. The
value of $L_{\rm sym}$, commonly considered to be close to 60\,MeV,
has been challenged by the recent analyzes of the neutral weak form
factor of $^{48}$Ca in the CREX experiment, which suggests a rather low
value of $L_{\rm sym} = 20 \pm 30$\,MeV~\citep{CREX:2022}. This can
be contrasted to the values of $L_{\rm sym} = 106 \pm 37$\,MeV
obtained by some groups via the analysis of the neutron skin thickness
of $^{208}$Pb in the PREX-II experiment~\citep{PREX-II:2021}.
Combined analysis of PREX and CERX experiments favors
$L_{\rm sym} \simeq 60$ MeV~\citep{Lattimer:2023}. Somewhat larger
value $L_{\rm sym} = 79 \pm 39$ MeV was deduced from collider-based
high-energy data involving the determination of the neutron skin of
$^{208}$Pb at LHC~\citep{Giacalone:2023}. The ab initio computations
predict $L_{\rm sym} = 37$-66\,MeV~\citep{Hu:2022NatPh}.
The lower range $L_{\rm sym}  = 20$ MeV used in our  modeling is
consistent with the value at which the CREX and PREX ranges begin
to overlap as $L_{\rm sym} $ is increased, see Figure~7 of~\cite{Lattimer:2023}. 

\begin{table}[tb]
\caption{Nuclear matter characteristics at saturation density.}
\setlength{\tabcolsep}{3.8pt}
\label{tab:Nuclear_matter}
\begin{tabular}{cccccccc}
\hline\hline
$\rho_0$   &$M^\ast_D$&$E_{\rm sat}$&$K_{\rm sat}$&$Q_{\rm sat}$&$J_{\rm sym}$&$L_{\rm sym}$\\
(fm$^{-3}$)& ($m_N$)  & (MeV)       & (MeV)       & (MeV)       & (MeV)       & (MeV)       \\
\hline
0.152      & 0.57     & $-16.14$    & 251.15      & $-600$ - 1000& 28.73 - 36.48& 20 - 100  \\
\hline
\end{tabular}
\tablecomments{$M^\ast_D$ denotes the Dirac effective mass in units of nucleon mass. 
The symmetry energy at the cross density 0.110 fm$^{-3}$ is fixed as 27.09 MeV the 
DDME2 value~\citep{Lijj:2023c}.}
\end{table}
\begin{figure*}[tb]
\centering
\includegraphics[width = 0.96\textwidth]{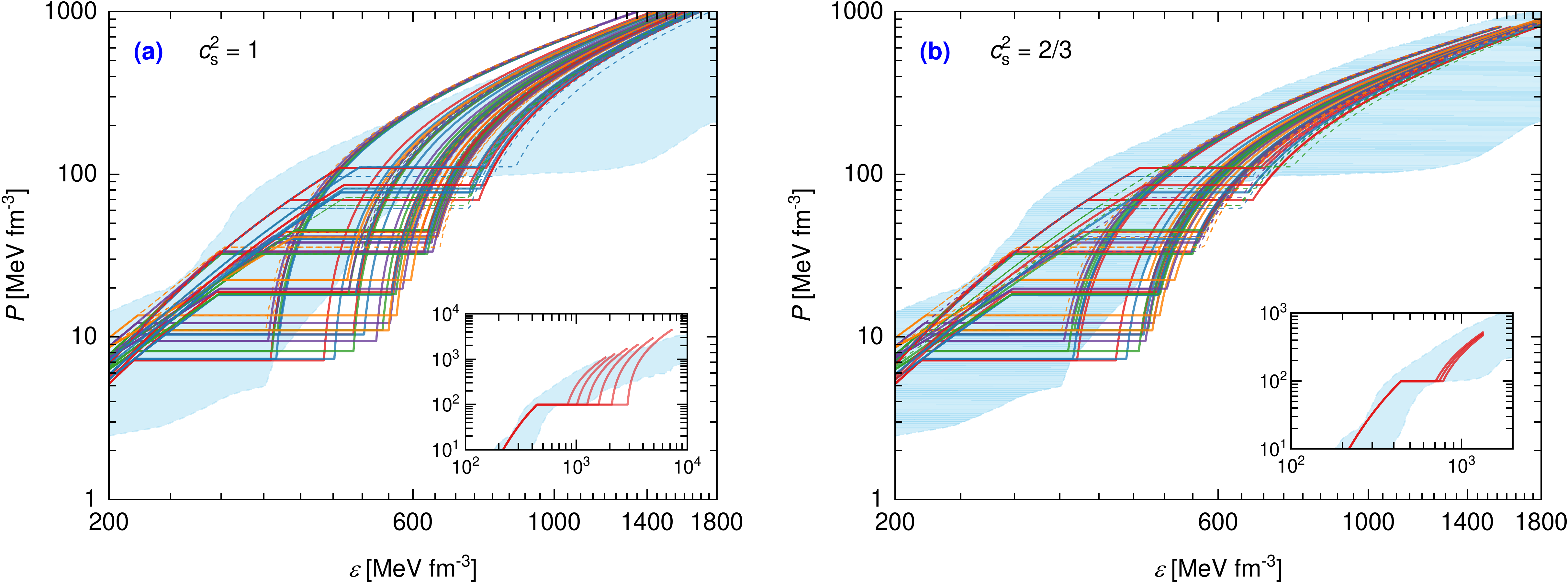}
\caption{ EoS collection utilized in the work.  Solid thick lines
  represent models that satisfy all constraints, while dashed thin
  lines show those that fail for HESS J1731-347. 
  The blue region shows the allowed range of EoS obtained by 
  imposing $M_{\rm max} \geq 2.0\,M_{\odot}$, $R_{2.0} \geq 11.1$\,km 
  and $\tilde{\Lambda}_{\rm GW170817} \leq 720$ according
  to~\cite{Annala:2022} (the yellow area of their Figure~1). 
  This includes EoSs mimicking first-order phase transitions. 
  Insets show EoSs that give rise to ultra-compact stars.
  }
\label{fig:EOS}
\end{figure*}

The quark phase is described by the constant sound speed (CSS) 
parameterization~\citep{Zdunik:2013,Alford:2013},
\bea\label{eq:EoS}
p(\ep) =\left\{
\begin{array}{ll}
p_{\rm tr}, \quad & \ep_{\rm tr} < \ep < \ep_{\rm tr}\!+\!\Delta\ep, \\[0.5ex]
p_{\rm tr} + c^2_s\,\bigl[\ep-(\ep_{\rm tr}\!+\!\Delta\ep)\bigr],
  \quad & \ep_{\rm tr}\!+\!\Delta\ep < \ep, 
\end{array}
\right.
\eea
where $p_{\rm tr}$ is the pressure and $\ep_{\rm tr}$ is the energy
density at the phase transition from nucleonic to quark matter at
zero temperature and nucleonic density $\rho_{\rm tr}$; $\Delta\ep$ is
the energy discontinuity and $c_s$ is the sound speed in the quark
phase. The first-order phase transition described by this EoS leads
to a new stable branch populated by hybrid stars that may or may not
be separated from the nucleonic branch by an instability region. 
Furthermore, for a range of parameters, this leads to the phenomenon 
of twin configurations --- two stable CSs that have identical masses 
but different radii and consequently, tidal
deformabilities~\citep{Alford:2013,Benic:2015,Paschalidis:2018,Alvarez-Castillo:2019,Christian:2019,Montana:2019,Christian:2021}.
In this work, we adopt three values of $c_s^2$: the maximally stiff EoS
with $c^2_s = 1$ (in natural units) allows us to explore the limits of
ranges of masses and radii accessible to models of the type considered
in this work; see also~\cite{Tsaloukidis:2023}. The alternative values
$c^2_s = 2/3$ and $1/2$ are intermediate between the maximally stiff and 
the asymptotic high-density value $c^2_s = 1/3$ corresponding to a conformally 
symmetric quark phase. Other values of $c^2_s$ have been considered 
very recently by several 
authors~\citep{Brodie:2023,Sagun:2023} in the context of HESS J1731-347
for alternate nucleonic EoSs.

\section{Computational Scheme}
\label{sec:Scheme}
To cover the parameter space of our model, we choose from the family
of 81 nucleonic EoS models given in \cite{Lijj:2023c} a subset which
are identified by the values of $Q_{\rm sat}$ and $L_{\rm sym}$. We
then choose the values for 
$\rho_{\rm tr} = 1.5,\,2.0,\,2.5,\,3.0\,\rho_0$ to match these nucleonic 
EoS to quark matter EoS, for a  magnitude of the density jump $\Delta\ep$ 
and squared sound-of-speed $c_s^2 = 1$, $2/3$ or $1/2$.

For each EoS in our collection, we solve the Tolman-Oppenheimer-Volkoff
(TOV) equations for static and spherically symmetric stellar objects. 
If all the mass and radius ranges of astronomical data are intersected 
by a sequence corresponding to a particular EoS it is accepted, otherwise 
it is rejected. More specifically, we consider the set of constraints
(a)-(c) listed in the Introduction, i.e., those based on NICER
inferences of mass and radius for J0030+0451 and J0740+6620, GW170817
inference of radii and masses in the binary CS merger, and radio-wave
measurements of the mass of J0740+6620. We then supplement these by
the mass and radius of J1731-347 and the mass of J0952-0607. Note that, 
if we assume uniform rotation, the increase of mass of the maximum-mass 
configuration due to the rotation of 707 Hz is about of $0.05\,M_{\odot}$. 
Therefore, the lower limit on the mass of J0952-0607 is about the same 
as that of J0740+6620 at a 95\% confidence level. The parameters for
quark EoS, $p_{\rm tr}$ and $\Delta\ep$, can be transformed into
observational properties of CSs, i.e., the maximum mass of the
nucleonic branch $M^{\rm H}_{\rm max}$ and that of the hybrid branch
$M^{\rm Q}_{\rm max}$, respectively, in the $M$-$R$
diagram~\citep{Lijj:2023a}. We further introduce the minimal mass in
the hybrid branch, denoted as $M^{\rm Q}_{\rm min}$, which we shall
use for assessing the range of mass where twin configurations exist.

\section{Analysis of the results}
\label{sec:Twins}
Figures~\ref{fig:MR_diagram1} and~\ref{fig:MR_diagram2} show the
$M$-$R$ diagrams for hybrid EoS models where the nucleonic EoS is
specified by values of $Q_{\rm sat}$ and $L_{\rm sym}$ and the quark
matter EoS is specified by the sound speed value $c^2_s = 1$, 2/3 and
1/2. Figure~\ref{fig:MR_diagram1} is for lower transition densities
$\rho_{\rm tr} = 1.5$ or $2.0\,\rho_0$; Figure~\ref{fig:MR_diagram2}
is for higher transition densities $\rho_\text{tr} = 2.5$ or
$3.0\,\rho_0$.  The isovector properties of nucleonic EoS are measured
by the value of $L_{\rm sym}$. Consider first the isovector-soft
(e.g., $L_{\rm sym} = 20$\,MeV) EoSs, which contain members that are
consistent with astrophysical constraints at the 95\% confidence level
for any of the chosen parameters shown in panels of
Figures~\ref{fig:MR_diagram1} and~\ref{fig:MR_diagram2}. It is seen
that as the transition density increases in the considered range
$1.5$-$3.0\,\rho_0$ the masses of twin stars increase as well.  A
larger speed of sound allows the hybrid branch to stretch to the
region of smaller radii, and more generally to cover a broader range,
as can be seen from the ultra-compact members shown in the insets in
panels (b, d) of Figure~\ref{fig:MR_diagram2}. These models can be
contrasted with the isovector-stiff (e.g., $L_{\rm sym}=80$\,MeV)
EoSs. In this case, the transition to quark matter needs to occur at a
sufficiently low density $\rho_{\rm tr} \lesssim 2.0\,\rho_0$, to
satisfy the constraints imposed by HESS J1731-347 and GW170817, as
displayed in Figure~\ref{fig:MR_diagram1}. Furthermore, compatibility
with the large radius of PSR J0740+6620 is achieved for relatively
small energy density discontinuity at the phase transition. The
resulting sequences satisfy the two-solar mass constraint because the
maximum of the hybrid branch $M^{\rm Q}_{\rm max}\sim 2.0\,M_{\odot}$,
which is also the maximum mass for the entire sequences. In this case,
only low-mass twin configurations appear that are consistent with
astrophysical constraints. 
For models with $c^2_s = 1/2$ twin configurations do not arise 
in the case of high-density transition from nucleonic to quark matter. 
They are present only for the lowest transition density 
$\rho_{\rm tr} = 1.5\,\rho_0$.

\begin{figure*}[tb]
\centering
\includegraphics[width = 0.98\textwidth]{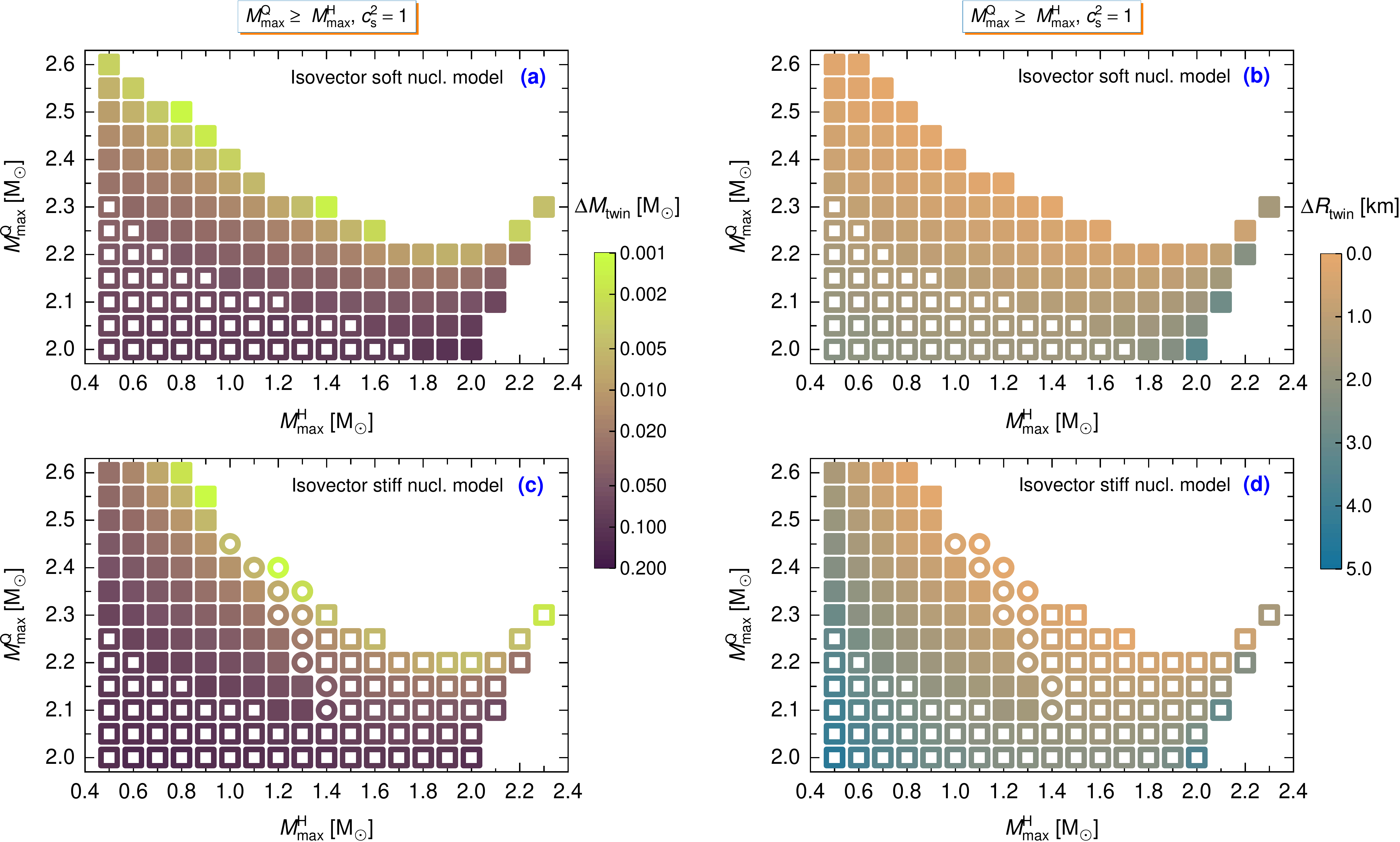}
\caption{ Ranges of mass and radius that characterize twin
  configurations for the case A, i.e., 
  $M^{\rm Q}_{\rm max} \geq M^{\rm H}_{\rm max}$. In panels (a, b)
  the hybrid EoS models ($c^2_s = 1$) are built using an
  isovector-soft nucleonic model
  ($Q_{\rm sat} = 800,\,L_{\rm sym} = 20$\,MeV), while in panels 
  (c, d) - using an isovector-stiff nucleonic model
  ($Q_{\rm sat} = 800,\,L_{\rm sym} = 80$\,MeV). Panels (a, c) show
  the mass ranges of twins $\Delta M_{\rm twin}$ using color
  coding. Panels (b, d) do the same but for the radii of twin
  configurations $\Delta R_{\rm twin}$.  By open squares, we show the
  models that are ruled out by combined observational constraints,
  whereas by open circles we show those that are excluded by HESS
  J1731-347 data. }
\label{fig:MM_relation}
\end{figure*}

Figure~\ref{fig:EOS} shows the high-density portion of all the EoSs
used in this work, i.e., including those that fail to account for HESS
J1731-347. 
(Models with $c^2_s = 1/2$ are not shown as they locate well
inside the range spanned by $c^2_s = 2/3$.)
The EoSs of ultra-compact stars, with radii in the range $6\le R \le 10$\,km 
and much higher central densities, are separated in the insets. It is seen
that the addition of J1731-347 does not change the range of allowed
EoS and that the high-density range covered by our collection is
broadly compatible with the one deduced in~\cite{Annala:2022} by
imposing current astrophysical constraints on a very large randomly
generated ensemble of physics-agnostic EoS covering the intermediate-density 
range. Some exception is seen among the EoS which feature low
transition density $\rho_{\rm tr}$ and large density jump $\Delta\ep$.
We also note that the low-density range of the EoS, which can be found
in~\cite{Lijj:2023c}, is consistent with the low-density microscopic
(chiral) EoS, as presented in~\cite{Annala:2022}, only for
$40\le L_{\rm sym}\le 60$\,MeV. 
Interestingly, despite the large variations of parameters of EoS, 
all models converge at high density to the region deduced by~\cite{Annala:2022}. 
This finding is contrary to their conclusion that perturbative QCD 
highly constrains nuclear matter properties at intermediate densities.

To facilitate the following discussion, we divided the stellar 
configurations into those where the maximum-mass star is on the hybrid 
branch (case A) and those where the maximum-mass star is on the nucleonic 
branch (case B). This type of classification has been considered previously; 
case A was studied in~\cite{Lijj:2021} and~\cite{Christian:2022} and 
case B in~\cite{Lijj:2023b}.

\begin{figure*}[tb]
\centering
\includegraphics[width = 0.98\textwidth]{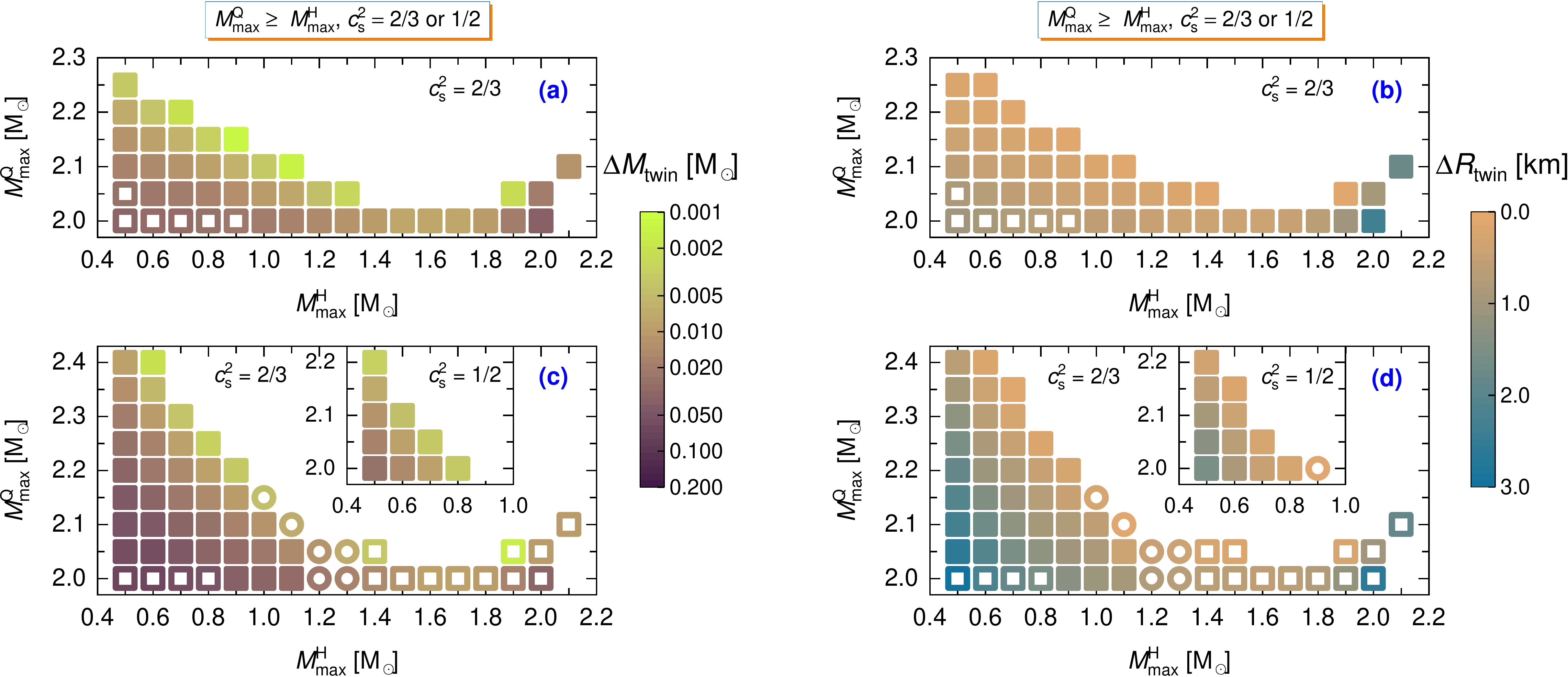}
\caption{ Same as in Figure~\ref{fig:MM_relation} but for models with 
intermediate values of $c^2_s = 2/3$ and $1/2$ (shown as insets). 
Low-mass twin configurations persist only for the isovector-stiff 
nucleonic model.
}
\label{fig:MM_relation2}
\end{figure*}

\underline{\it Case A:} Figure~\ref{fig:MM_relation} summarizes the 
contours representing the ranges of mass and radius that characterize 
twin configurations for hybrid EoS models with 
$M^{\rm H}_{\rm max} \leqslant M^{\rm Q}_{\rm max}$ for $c_s^2 = 1$. 
Filled symbols indicate models that satisfy astrophysical constraints, 
while empty symbols represent models that do not. The color of the square 
shows the range of mass or radius over which twins can be found. 
The cases $c_s^2=2/3$ and $1/2$ in Figure~\ref{fig:MM_relation2} show 
the same trends but the ranges of masses and radii are significantly narrower. 
The shape of the sequences in Case A dictates that the range of masses where 
twin configurations exist is defined as 
$\Delta M_{\rm twin}=M^{\rm H}_{\rm max}-M^{\rm Q}_{\rm min}$. 
Then, the differences in radii $\Delta R_{\rm twin}$ for twin configurations
is determined by comparing the radius of $M^{\rm H}_{\rm max}$
nucleonic star to that of the hybrid counterpart with an identical
mass. It can be observed from Figures~\ref{fig:MR_diagram1} and
\ref{fig:MR_diagram2} that this is also the maximal radius difference for 
a hybrid EoS predicting twin configurations. A lesson that we learn from 
Figures~\ref{fig:MM_relation} and~\ref{fig:MM_relation2} is that the possible 
combinations $M^{\rm Q}_{\rm max}$ and $M^{\rm H}_{\rm max}$ for which
twin configurations exist is large for the isovector-soft nucleonic model
[(panel (a)] and is small for the isovector-stiff model [(panel (c)]. 
Also these two cases predict different $M^{\rm Q}_{\rm max}$ and
$M^{\rm H}_{\rm max}$ for which the largest range of masses
$\le 0.2\,M_{\odot}$ (shown by dark squares) is achieved. The
corresponding values of $\Delta R_{\rm twin}$ shown in panels (b, d) 
indicate that the largest difference between the twin radii is
achieved for sequences that have both maxima around the two-solar mass.
The twin configurations with $\Delta R_{\rm twin} \sim 4$--5\,km correspond 
to low-mass hybrid stars relevant for HESS J1731-347.

\begin{figure}[b]
\centering
\includegraphics[width = 0.45\textwidth]{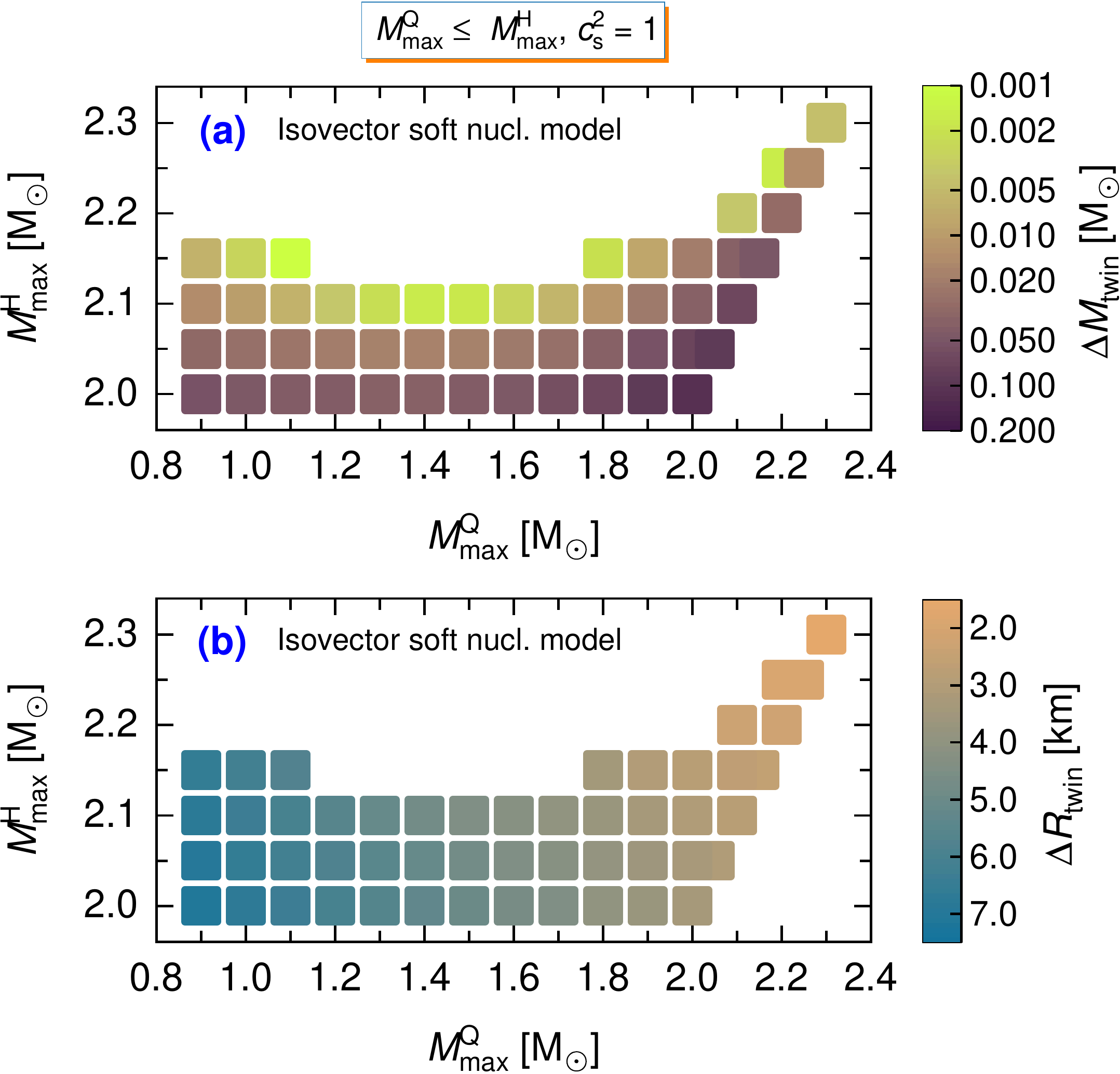}
\caption{ Same as in Figure~\ref{fig:MM_relation} but for the case B 
  corresponding to $M^{\rm Q}_{\rm max} \leq M^{\rm H}_{\rm max}$. 
  The hybrid EoS models ($c^2_s = 1$) are built upon an isovector-soft 
  nucleonic model ($Q_{\rm sat} = 800,\,L_{\rm sym} = 20$\,MeV). 
  Ultra-compact configurations having large $\Delta R_{\rm twin}$ are 
  present only for models with 
  $M^{\rm H}_{\rm max} \simeq 2.0$-$2.15\,M_{\odot}$.
}
\label{fig:MM_ultra}
\end{figure}

\underline{\it Case B:} Figure~\ref{fig:MM_ultra} summarizes the 
ranges of masses and radii that characterize twin configurations 
for hybrid EoS models in the case 
$M^{\rm Q}_{\rm max} \leqslant M^{\rm H}_{\rm max}$, 
for $c_s^2 = 1$. Here the range of masses is defined
$\Delta M_{\rm twin}= M^{\rm Q}_{\rm max}-M^{\rm Q}_{\rm min}$ and 
the differences in radii $\Delta R_{\rm twin}$ is extracted from the
radius of $M^{\rm Q}_{\rm max}$ hybrid star and its nucleonic twin.
For such cases, as illustrated in the insets of 
Figure~\ref{fig:MR_diagram2}, we require the nucleonic branch passing
through all the 95\% confidence regions of constraints, 
this can be achieved only for isovector-soft models and the values 
of $c^2_s$ close but below unity~\citep{Lijj:2023b}. 
It is seen that ultra-compact configurations with radii 
$6\le R\le 10$\,km, associated with large values of 
$\Delta R_{\rm twin}$, are present only for models with
$M^{\rm H}_{\rm max} \simeq 2.0$-$2.15\,M_{\odot}$.
These ultra-compact members studied in more detail in~\cite{Lijj:2023b} 
together with those appearing in case A offer an alternative to 
strange stars, which up until recently were considered as the only 
possible objects with small radii 
(of the order of $6$--$9$\,km)~\citep[see][and references therein]{Bombaci:2021}. 
It is worthwhile also to mention that the maximum-mass values found for 
hybrid EoSs featuring twin configurations in this case are well within 
the range of mass obtained for PSR J0592-0607 at a 95\% confidence 
level~\citep{Romani:2022}.

%

\section{Conclusions}
\label{sec:Conclusions}
In this work, we investigated the consistency of the hybrid star
models with the current astrophysical and laboratory data and explored
the range of parameters that lead to twin stars. To do so, we employed
a recently constructed family of nucleonic EoSs parameterized by
the $Q_{\rm sat}$ and $L_{\rm sym}$ parameters of the Taylor expansion
of the energy density of nucleonic matter. This was then combined with
a CSS parameterization of the quark matter EoS for a range of values
of the transition density, energy density jump, and sound speed.
We have identified two classes of models with low and high transition
densities (as shown in Figures~\ref{fig:MR_diagram1} and~\ref{fig:MR_diagram2},
respectively) that are consistent with the current data. This is
achieved through a compromise between the requirement for a soft
EoS at low densities, driven by the data from HESS J1731-347 and GW170817, 
and the need for stiff EoS at high densities to account for the 
radio-astronomy data for PSR J0592-0607 and J0740+6620. The ranges of 
parameters of twin configurations that are allowed by our hybrid models 
were found as shown in Figures~\ref{fig:MM_relation} and~\ref{fig:MM_ultra}. 
Notably, the data from J1731-347 supports the existence of low-mass twin 
configurations with masses $M \lesssim 1.3\,M_{\odot}$, regardless of 
the chosen nucleonic EoS, but it requires a low value for the transition 
density. These hybrid models also allow for maximum masses 
$M^{\rm Q}_{\rm max}$ ranging from $2.0$ to $2.6\,M_{\odot}$ depending on 
the value of the sound speed in quark matter. In addition, ultra-compact 
configurations, which lie outside the constraints imposed by GW170817 
and J1731-347, emerge for isoscalar-stiff nucleonic EoS models and for
large speed-of-sound $c^2_s=1$ in the case where the nucleonic branch
extends up to $2.0$--$2.15\,M_{\odot}$.

In conclusion, the current astrophysical and nuclear physics data do
not rule out hybrid stars, and leave room for hybrid-nucleonic mass
twins if a strong first-order phase transition occurs in dense matter
for reasonable values of nucleonic isoscalar and isovector
characteristics. It is even plausible to argue that all the stars with
observational information we have considered in this study are 
hybrid configurations. But this does not exclude the possibility
that current observational data are accounted for by the nucleonic
branch of hybrid sequences, whereas the hybrid branch lies outside
of ranges currently covered, as would be the case for ultra-compact 
members of sequences.

\section*{Acknowledgements}
J.~L. acknowledges the support of the National Natural Science 
Foundation of China (Grant No. 12105232), the Fundamental Research 
Funds for the Central Universities (Grant No. SWU-020021), and by 
the Venture \& Innovation Support Program for Chongqing Overseas 
Returnees (Grant No. CX2021007). A.~S. acknowledges the DFG Grant
No. SE 1836/5-2 and the Polish NCN Grant No. 2020/37/B/ST9/01937 
at Wroc\l{}aw University. M.~A. is partly supported by the U.S. 
Department of Energy, Office of Science, Office of Nuclear Physics 
under Award No. DE-FG02-05ER41375.

\bibliographystyle{aasjournal}
\bibliography{Twins_refs}
\end{document}